\documentclass[letterpaper, 11 pt, conference]{article}  
\usepackage{cite}
\usepackage{amsfonts}
\usepackage[ruled,vlined]{algorithm2e}
\usepackage{textcomp}
\usepackage{float}
\usepackage{graphicx}
\usepackage{bm}
\usepackage{amsmath}
\usepackage{amssymb}
\usepackage{hyperref}
\usepackage{makecell}
\usepackage{xfrac}
\usepackage[margin=0.8in]{geometry}

\usepackage{color}
\usepackage{subcaption}
\usepackage{lipsum} 

\begin{document}

\title{\bf Policy Gradient-Based EMT-in-the-Loop Learning to Mitigate Sub-Synchronous Control Interactions }

\author{Sayak Mukherjee$^\dagger$, Ramij R. Hossain$^\dagger$, Kaustav Chatterjee$^\dagger$, \\ Sameer Nekkalapu$^\dagger$, Marcelo Elizondo
\thanks{Authors are with the Pacific Northwest National Laboratory, 902 Battelle Blvd, Richland WA 99354, $^\dagger$ contributed equally. The research is supported by the E-COMP (Energy System Co-Design with Multiple Objectives and Power Electronics) Initiative at Pacific Northwest National Laboratory (PNNL), operated by Battelle for the U.S. Department of Energy under Contract DEAC05-76RL01830.}%
}
\date{}
\maketitle

\begin{abstract}
This paper explores the development of learning-based tunable control gains using EMT-in-the-loop simulation framework (e.g., PSCAD interfaced with Python-based learning modules) to address critical sub-synchronous oscillations. Since sub-synchronous control interactions (SSCI) arises from the mis-tuning of control gains under specific grid configurations, effective mitigation strategies require adaptive re-tuning of these gains. Such adaptiveness can be achieved by employing a closed-loop, learning-based framework that considers the grid conditions responsible for such sub-synchronous oscillations. This paper addresses this need by adopting methodologies inspired by Markov decision process (MDP) based reinforcement learning (RL), with a particular emphasis on a simpler deep policy gradient methods with additional SSCI-specific signal processing modules such as down-sampling, bandpass filtering, and oscillation energy dependent reward computations. Our experimentation in a real-world event setting demonstrates that the deep policy gradient based trained policy can adaptively compute gain settings in response to varying grid conditions and optimally suppress control interaction-induced oscillations.

\end{abstract}

\section{Introduction}

Sub-synchronous control interactions (SSCIs) are sustained oscillations resulting from adverse couplings between the fast inner/outer control loops of inverter-based resources (IBRs) and the network at frequencies below or above the nominal synchronous frequency \cite{stability_def, cheng2022real}. Historically linked to wind farms connected through series-compensated transmission lines, they are now also observed in uncompensated or weak-grid conditions \cite{cheng2022real}—for example, 3.5 Hz real/reactive power oscillations in Hydro One’s network \cite{li2019asset}, 4 Hz voltage-control oscillations in ERCOT’s Texas grid \cite{Texassystem_Mainpaper}. Such interactions can accelerate equipment degradation, cause protection misoperations, and threaten overall system stability, motivating mitigation via converter retuning, damping control, and coordinated control designs. Series-compensator SSCI arises from interactions between IBR control dynamics and the resonant modes of series-compensated transmission lines. Moreover, weak-grid SSCIs stem from control and grid impedance interactions in systems with low short-circuit strength \cite{fan2022real}.

 Consequently, increasing research efforts have focused on developing control solutions to mitigate SSIs/SSCIs in inverter-rich grids. Prior work has proposed damping strategies for doubly-fed induction generators (DFIGs) and series-compensated transmission systems \cite{abdeen2022sub}, as well as robust coordinated schemes for suppressing sub-synchronous oscillations \cite{wang2020robust}. Adaptive multi-modal damping controllers tailored for weak grids have also been reported \cite{shair2023grid}. 
 Data-driven and machine learning–based methods for SSI mitigation are presented in \cite{liu2024mitigating}. Deep reinforcement learning has been utilized for supplementary damping control \cite{liu2025sub}, emergency action based decision making \cite{zhu2025emergency}, inverter tuning using Simulink-based dynamic link library \cite{das2025deep} for SSO mitigation. 

\noindent \textit{Research Gap and Contributions.} However, since the SSCI phenomenon often arises from mis-tuned control gains under specific configurations of surrounding grid elements, a practical mitigation strategy should emphasize adaptive gain re-tuning. Such adaptiveness can be achieved through a closed-loop learning-based framework incorporating high-fidelity EMT simulators such as PSCAD, and that explicitly accounts for the grid conditions responsible for exciting sub-synchronous oscillation modes, with an emphasis on testing in a real-world event. This paper focuses on developing such framework using Markov decision process (MDP)–based reinforcement learning (RL) methodologies \cite{sutton1998introduction, mnih2015human}, and in particular from policy-gradient methods \cite{sutton1999policy, schulman2015trpo, Schulman2017}. In contrast to prior studies that primarily employ complex actor–critic or off-policy DRL architectures, this work demonstrates a simpler deep policy-gradient framework directly embedded in an EMT-in-the-loop PSCAD environment. The proposed approach develops tunable control gains using an EMT-in-the-loop simulation setup to actively mitigate critical sub-synchronous oscillations. The implementation integrates PSCAD with an outer-loop Python API that invokes the simulator and dynamically adjusts controller parameters during runtime. A dedicated data-processing routine performs down-sampling and band-pass filtering to construct the observation windows used for learning. The control policy is modeled as a multi-layer perceptron (MLP)–based deep neural network, while the reward function is formulated in terms of the oscillation energy computed from the inverter’s active-power response. Once trained, the deep policy-gradient agent provides an adaptive gain computation mechanism that continuously re-tunes the controller in response to changing grid conditions, thereby achieving optimal damping of control-interaction-induced oscillations.

\section{Learning Methodology}
\subsection{Framework Overview}
\begin{figure}[t!]
    \centering
    \includegraphics[width=0.7\linewidth]{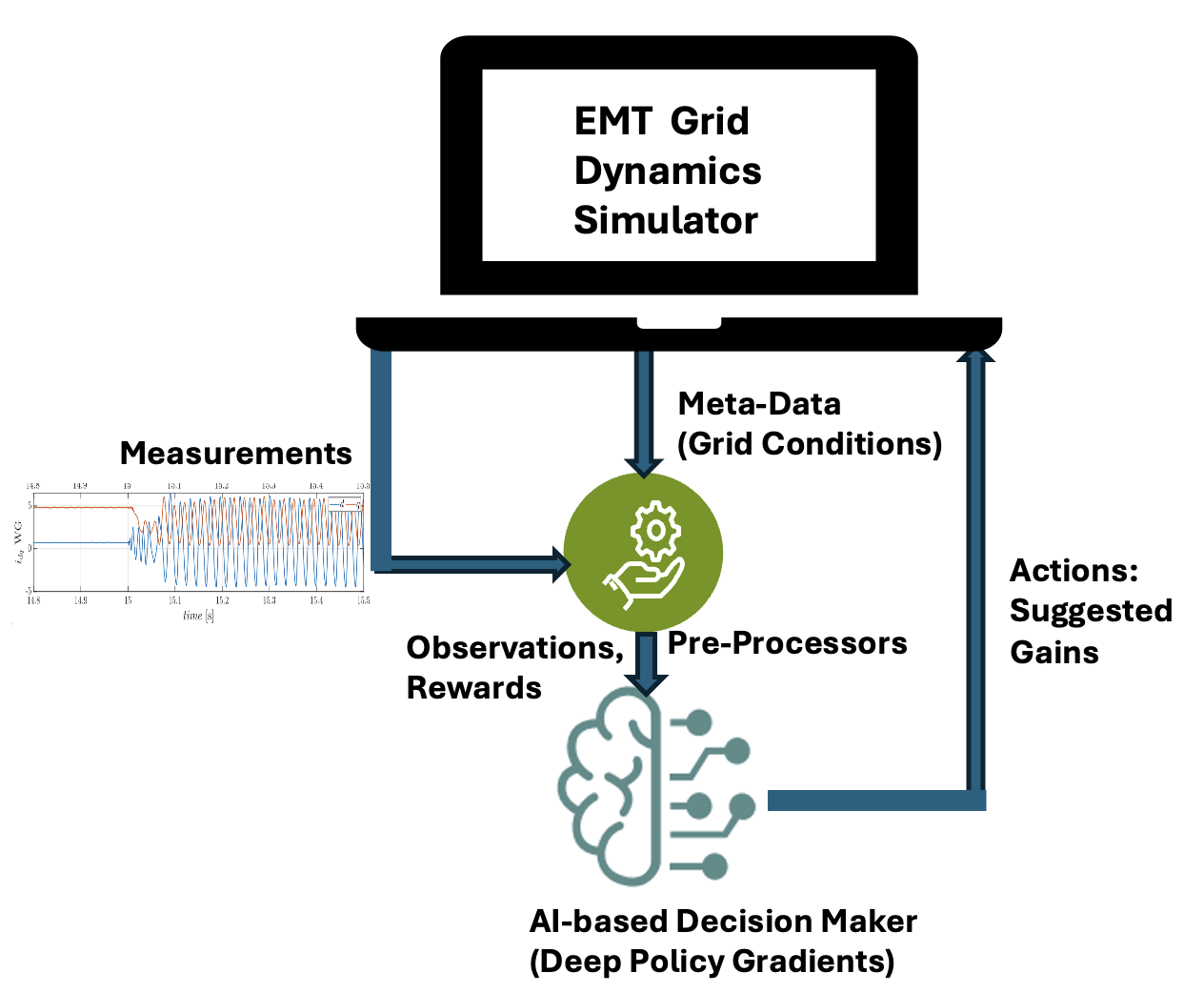}
    \caption{Schematic of the AI-based Mitigation of SSCI with EMT-in-th-loop Training}
    \label{fig:overview}
\end{figure}

This research investigates learning tunable gains of the outer and inner control loops of the IBR that induces SSCI with EMT-in-the-loop simulators to mitigate critical oscillations. We consider the Texas grid example, and consider a SSCI event, where the gains of the wind farm controller needs to be properly tuned in order to mitigate any adverse oscillations. In series-compensator SSCIs, instability occurs when the network resonance coincides with inverter control bandwidths (e.g., of active/reactive power or voltage regulators). Excessive PI gains and phase-locked loop (PLL) interactions can introduce negative damping, allowing transient energy circulation among control loops, the PLL, and the series-compensated network \cite{cheng2022real}. Fig. \ref{fig:overview} shows the overall architecture for the AI-based closed-loop control design to mitigate SSCIs. The goal of the learning control agent is to counteract SSCI induced oscillations. Given the complex nature of oscillations, rule-based designs are insufficient, therefore, we formulate this problem as a (partially observable) Markov Decision Process (MDP), defined by the tuple $(S, A, \mathcal{P}, r, \gamma)$ \cite{sutton1998introduction}, where the state space (representing grid dynamics) $S \subseteq \mathbb{R}^n$ and the action space (tunable gains of outer and inner control loops of IBRs) $ A \subseteq \mathbb{R}^m$ are continuous. The environment transition function $\mathcal{P}: S \times A \to S$ characterizes the stochastic transition of the grid states during dynamic events, and the reward function $r: S \times A \to R$ along with the discount factor $\gamma \in (0,1)$ are also defined. For the development of this framework, we consider the following design considerations.

  \noindent  \textit{EMT-in-the-loop Training Setup:} We developed a computational co-simulation framework that integrates PSCAD, an electromagnetic transient (EMT) simulator, with an external Python-based interface capable of bidirectional communication. Using the PSCAD Application Programming Interface (API), the framework enables Python modules to send real-time control commands (e.g., tunable control gains) and retrieve system measurements directly from the simulator. This design allows the external training or control algorithm to interact dynamically with the EMT environment, facilitating closed-loop testing, adaptive parameter tuning, and data-driven controller evaluation.

   \noindent \textit{Data extraction module:} The data extraction script performs required down-sampling when needed, and band-pass filtering to remove any offsets that may be present in the raw measurements. These essential processing scripts help curate the desired observation windows, that will be passed to the AI-based decision maker to perform reward computation and utilized as feedback measurements.
Let the raw measured active power signals from the PSCAD--Python interface be denoted as $
P_{\text{meas}}(t), t \in [0, T],
$
where $T$ is the total simulation horizon.

\noindent\textit{1) Down-sampling:}
The raw data are down-sampled from the EMT sampling rate $f_s$ to a lower rate $f_s'$ using a down-sampling operator $\mathcal{D}_{f_s \to f_s'}(\cdot)$:
\[
P_d(k) = \mathcal{D}_{f_s \to f_s'}\!\big(P_{\text{meas}}(t)\big), 
\quad k = 1, \dots, N.
\]

\noindent\textit{2) Band-pass filtering:}
To eliminate DC offsets and unwanted high-frequency noise, a band-pass filter $\mathcal{F}_{[f_{\min}, f_{\max}]}(\cdot)$ is applied:
\[
P_f(k) = \mathcal{F}_{[f_{\min}, f_{\max}]}\!\big(P_d(k)\big).
\]

\begin{algorithm}[t!]
    \SetAlgoLined
    \caption{Policy Gradient Training with PSCAD-in-the-Loop}
    \label{alg:PG_PSCAD}
    \DontPrintSemicolon
    {\bf Inputs:} Number of epochs $n_\text{epoch}$, iterations per epoch $n_\text{iter}$, policy $\pi_\theta$, learning rate $\alpha$, optimizer.\;
    Initialize best reward $R^* \gets -\infty$.\;
    
    \For{$e = 1$ to $n_{\text{epoch}}$}{
        Initialize buffers for log-probabilities, actions, and rewards.\;
        
        \For{$n = 1$ to $n_{\text{iter}}$}{
            Obtain PSCAD measurements:
            \[
                P_{\text{meas}}(t), \quad t \in [0, T].
            \]
            
            Construct processed observation:
            \[
                s = \mathcal{W}_{T_w}\!\Big(
                        \mathcal{F}_{[f_{\min}, f_{\max}]}\!
                        \big(
                            \mathcal{D}_{f_s \to f_s'}(P_{\text{meas}})
                        \big)
                    \Big).
            \]
            
            Sample action from policy:
            \[
                a \sim \pi_\theta(\cdot \mid s), 
                \qquad \ell = \log \pi_\theta(a \mid s).
            \]
            
            Apply $a$ to PSCAD and obtain the system response:
            \begin{align}
P_{\text{meas}}(t+\Delta t) = f(s(t+\Delta t)), \\
s(t+\Delta t)= \mathcal{G}_{\text{PSCAD}}\!\big( s(t), \, a_i \big).
\end{align}
            
            Compute reward:
            \[
                R = -\!\int_{0}^{T_w}\!(P_f(t) - P_{\text{nom}})^2 dt.
            \]
            
            Store $(\ell, R)$ for gradient update.\;
        }
        
        Compute policy gradient objective:
        \[
            J(\theta) = 
            \frac{1}{n_{\text{iter}}}
            \sum_{j=1}^{n_{\text{iter}}} 
            \ell_j \, R_j.
        \]
        
        Update policy parameters:
        \[
            \theta \leftarrow 
            \theta + \alpha \nabla_\theta J(\theta).
        \]
        
        Record average epoch reward and update best model if improved.\;
    }
    
    {\bf Return:} Optimized policy parameters $\theta^*$. \;
\end{algorithm} 
\noindent\textit{3) Observation window extraction:}
Observation segments of duration $T_w$ are then constructed using a windowing operator $\mathcal{W}_{T_w}(\cdot)$:
\[
o_i = \mathcal{W}_{T_w}\!\big(P_f\big), 
\quad i = 1, \dots, M,
\]
where each $o_i \in \mathbb{R}^{T_w}$ represents the observation fed to the learning agent. For simplicity, we use the notation $o \in \mathbb{R}^{d_{\text{obs}}}, d_{\text{obs}} = T_w$, for subsequent discussion, however, during training we randomized this based on different windowing operations.

\noindent \textit{Reward Design:} Reward ($R$) is defined using the computed oscillation energy under the active power generated by the WF-IBR such as:
    \begin{equation}
E_{\text{osc}} \;=\; \int_{0}^{T_w} \big( P_f(\tau) - P_{\text{nom}} \big)^{2} \,
d\tau,\\
R = -E_{\text{osc}},
\end{equation}
for the observation window captured by final time $T_w$. When a band-pass filtered active power output signal is utilized we will have, $P_{nom} = 0$, and the signal itself will contain all the effect caused due to the oscillations. 

\noindent \textit{Policy:} 
We parameterize the policy as a Gaussian distribution:
\begin{equation}
    \pi_\theta(a \mid o) = \mathcal{N}\big(\mu_\theta(o), \, \sigma^2_\theta(o)\big),
\end{equation}
where $o \in \mathbb{R}^{d_{\text{obs}}}$ is the observation and the mean and variance are obtained from a neural network:
\begin{align}
    h_1 &= \mathrm{ReLU}\!\big(W_1 \,\mathrm{LN}(o)\big), \\
    h_2 &= \mathrm{ReLU}\!\big(W_2 h_1\big), \\
    \mu_\theta(o) &= W_3 h_2, \\
    \sigma^2_\theta(o) &= \mathrm{Softplus}(W_3' h_2).
\end{align}

The controller samples actions using the reparameterization trick:
\begin{equation}
    a = \mu_\theta(o) + \sigma_\theta(o) \odot \epsilon, 
    \qquad \epsilon \sim \mathcal{N}(0, I).
\end{equation}

The log-probability of the sampled action is
\begin{equation}
    \log \pi_\theta(a \mid o) 
    = -\tfrac{1}{2}\Bigg[ 
        \frac{(a - \mu_\theta(o))^2}{\sigma^2_\theta(o)} 
        + \log\!\big(2\pi \sigma^2_\theta(o)\big)
    \Bigg],
\end{equation}

We train the policy using an episodic policy gradient method with PSCAD-in-the-loop simulation as described in Algorithm \ref{alg:PG_PSCAD}. To decrease the EMT computational burden, we employ a restricted training strategy where if the suggested gains from the AI agent are within a preset range we just use a representative PSCAD simulation run to evaluate its dynamic performance, and restrict the action space. 
The overall PSCAD--Python framework operates as a continuous feedback loop:
\[
\begin{aligned}
P_{\text{meas}}(t) 
&\xrightarrow{\;\mathcal{D},\,\mathcal{F},\,\mathcal{W}\;} 
o
\xrightarrow{\;\pi_\theta\;} 
a
\xrightarrow{\;\text{PSCAD}\;} 
P_{\text{meas}}(t + \Delta t),
\end{aligned}
\]
which can be compactly expressed as the closed-loop update:
\begin{align}
P_{\text{meas}}(t+\Delta t) = f(s(t+\Delta t)), \\
s(t+\Delta t)= \mathcal{G}_{\text{PSCAD}}\!\big( s(t), \, a_i \big),
\end{align}
where $\mathcal{G}_{\text{PSCAD}}(\cdot)$ represents the nonlinear EMT system response simulated by PSCAD under the applied control command, with the grid states are denoted by $s(t)$, and the observations are a nonlinear function $f(.)$ of states.

\vspace{0.3 cm}

\noindent \textit{Restricted Action Space Training with Limited Experiments:} 
 To reduce the computational burden associated with electromagnetic transient (EMT) simulations, which are inherently time-consuming due to their fine time-step resolution and detailed switching dynamics, we adopt a restricted training strategy for the AI controller. In this approach, the learning agent is not allowed to explore the entire continuous action space freely. Instead, when the controller’s proposed gain values fall within a predefined \textit{safe}  range, we bypass the need to run a full EMT simulation for every trial with a representative trajectory information.

Rather than executing numerous EMT runs for small variations in similar gain settings, we use a representative EMT simulation corresponding to that region of the action space to approximate the system’s dynamic performance. This effectively clusters similar control actions into regions evaluated by a single high-fidelity PSCAD simulation, while constraining further exploration around that local neighborhood. By doing so, the AI agent still receives meaningful performance feedback but avoids redundant EMT computations. This significantly reduces the total simulation time during training, while ensuring that only dynamically distinct control configurations trigger new EMT evaluations.

\section{Test System Based on a Real-World Event and Experimental Results}
In 2009, the southern Texas grid experienced a sub-synchronous control interaction (SSCI) involving a DFIG-based wind farm \cite{Texassystem_Mainpaper, Texassystem_Suggestedpaper}. After a 345 kV line tripped, the plant was radially connected through a 50 \% series-compensated line, where interactions between DFIG controls and the capacitor resonance induced strong sub-synchronous oscillations, raising major stability concerns.
\begin{figure}[t!]
    \centering 
    \includegraphics[width=0.8\linewidth]{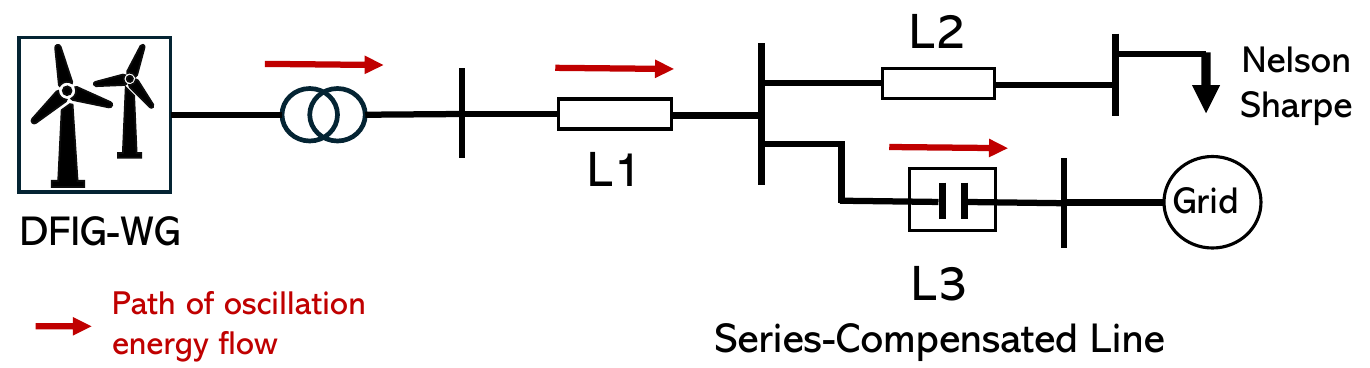}
    \caption{Single-line diagram of the test system, used to replicate the SSCI event from Texas, showing the path of oscillation energy flow or control interaction.} 
    \label{fig:Circuit_Texas}
    \vspace{0.3 cm}\includegraphics[width=0.8\linewidth]{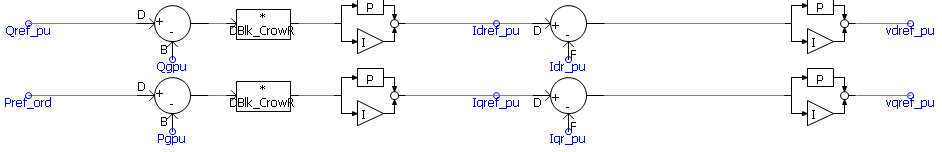}
    \caption{Outer and Inner Control Loops of the WF IBR}
    \label{fig:dfig_control}
\vspace{-0.3 cm}
\end{figure}
\begin{figure}[t!]
    \centering \vspace{-0.2cm}
    \includegraphics[width=0.7\linewidth]{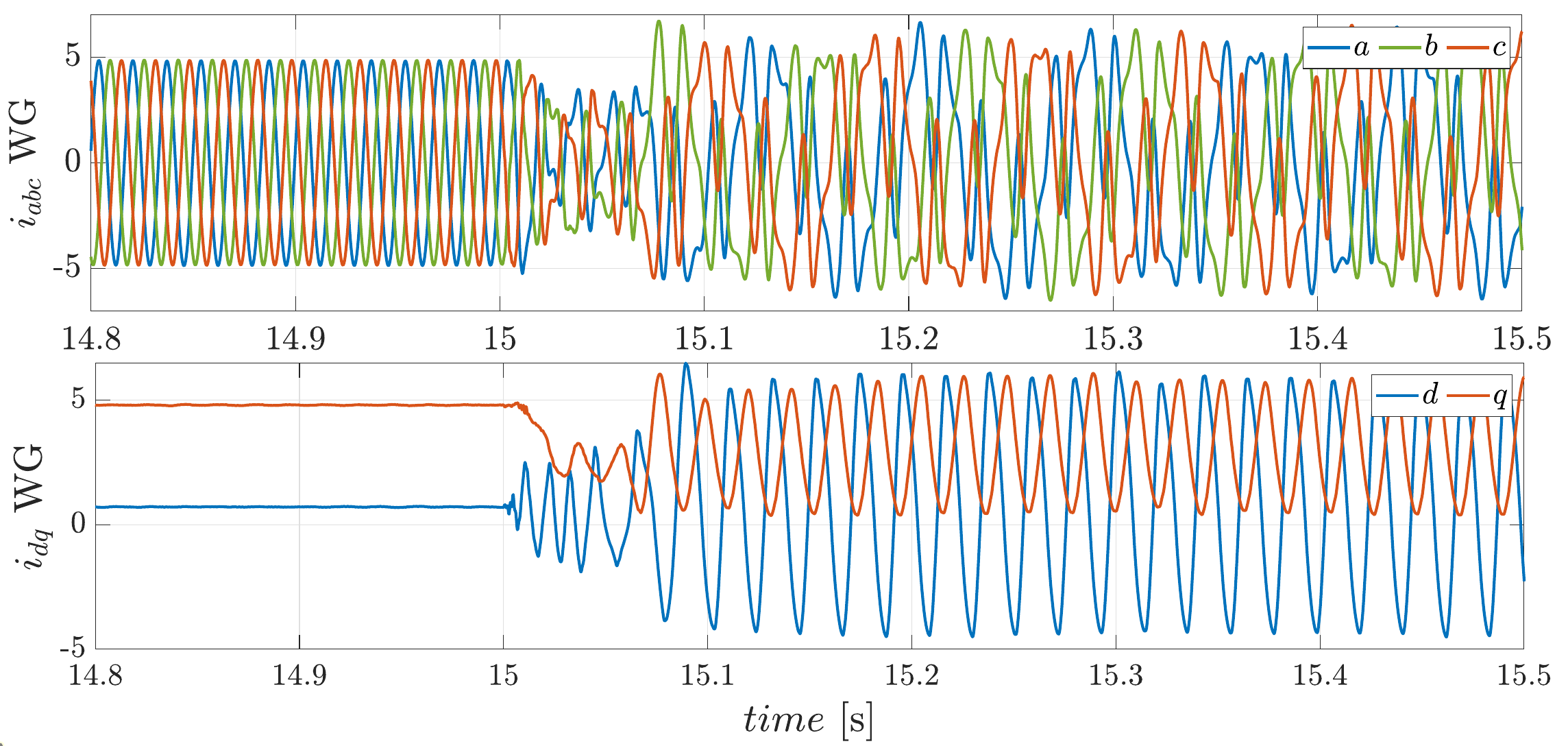} \vspace{-0.5cm}
    \caption{$abc$ phase waveforms and $dq$ components of the current injected by the WG in the Texas system.}
    \label{fig:Currents_WG_Texas}
    \centering\vspace{-0.1cm}
    \includegraphics[width=0.7\linewidth]{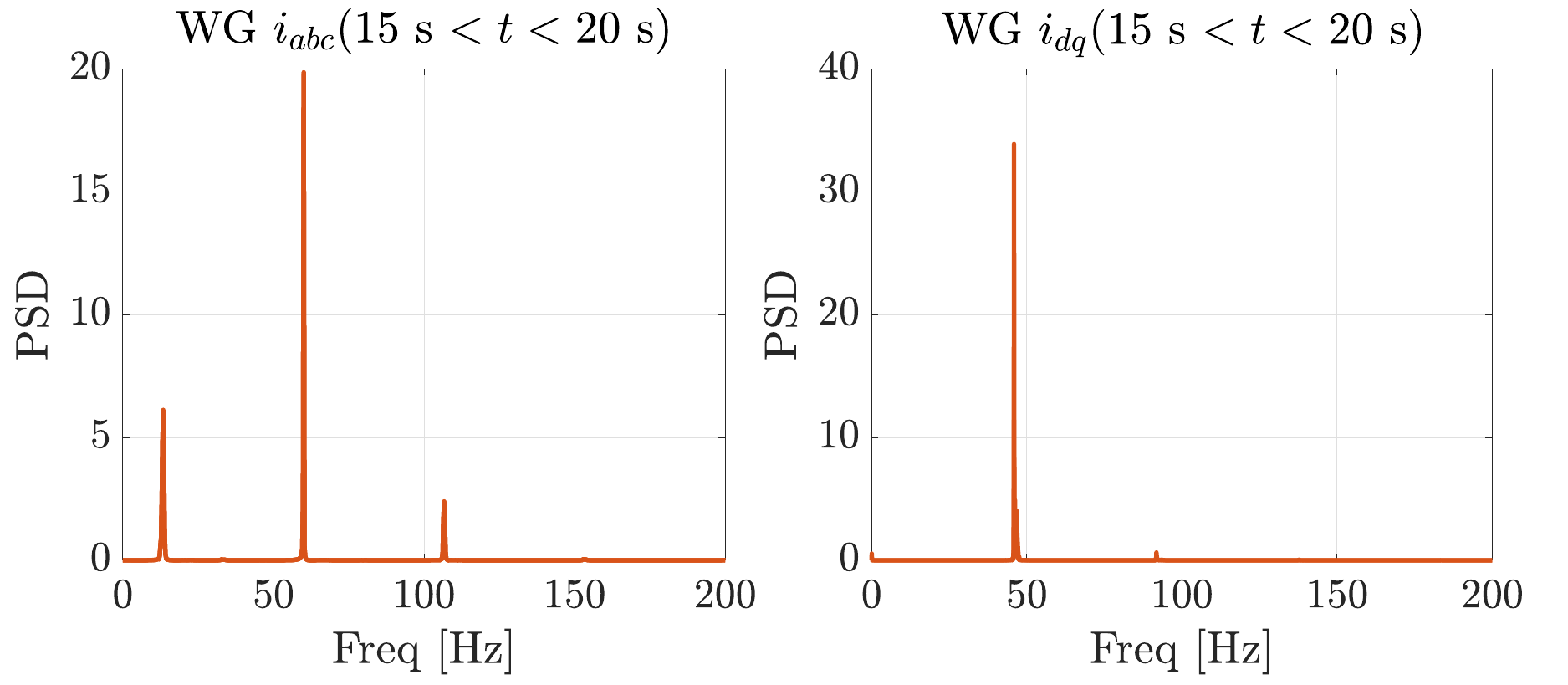}
    \caption{Frequency spectrum of the currents injected by the WG computed for (a) instantaneous three-phase components and (b) $dq$ components. }
    \label{fig:psd_WG_Texas}
\end{figure}
The event was replicated in the EMT platform PSCAD \cite{chatterjee2025identification} using the network illustrated in Fig.~\ref{fig:Circuit_Texas}. The DFIG-based wind plant delivers about 200 MW of power, while the Nelson Sharpe bus load is set to be negligible to represent post-fault conditions following the 345 kV line outage. The series-compensation capacitance is 400 µF, identified through sensitivity studies as the lowest value ensuring stability under nominal operation. A parallel breaker enables soft capacitor switching to mitigate overvoltages. Transmission lines L1–L3, linking the Zorillo–Ajio, Ajio–Nelson Sharpe, and Ajio–Rio Hondo buses, are modeled as lumped 60 Hz $\pi$-equivalents. In the simulated scenario (Fig.~\ref{fig:Currents_WG_Texas}), the proportional gains \(K_p\) of the outer rotor-side control loops of the DFIG (Fig. \ref{fig:dfig_control}) are increased from 2 to 4 at \(t = 15~\text{s}\). This adjustment excites sideband components at 12~Hz and 108~Hz, along with their harmonics, in the instantaneous phase currents due to the DFIG–series-capacitor interaction. In the \(dq\)-frame, these components manifest as a shifted frequency of 48~Hz (\(60 - 12 = 48~\text{Hz}\) or \(108 - 60 = 48~\text{Hz}\)), as illustrated in Fig.~\ref{fig:psd_WG_Texas}. The instability stems from the coupling between the DFIG outer-loop control and the network resonance introduced by the series-compensated line. The oscillatory energy circulates among the converter control loops, the inner current controller, and the resonant network branch through lines L1 and L3 (Fig.~\ref{fig:Circuit_Texas}), thereby sustaining the observed sub-synchronous oscillations.

\begin{table*}[t!]
\centering
\small
\caption{Training Parameters and Hyperparameters for Policy Network}
\begin{tabular}{l l l}
\hline
\textbf{Parameter} & \textbf{Symbol / Variable} & \textbf{Value / Description} \\
\hline
Observation dimension & \texttt{obs\_dim} & 30 (after the windowing operation) \\
Action dimension & \texttt{action\_dim} & 1 (all the outer-loop $K_p$ gains are tuned jointly) \\
Hidden layer size & \texttt{hidden\_size} & 64 \\
Learning rate & \texttt{lr} & $1 \times 10^{-3}$ \\
Optimizer & \texttt{optimizer} & Adam ($\beta_1=0.9$, $\beta_2=0.999$) \\
Activation function & -- & ReLU \\
Normalization & \texttt{LayerNorm} & Applied to input layer \\
Network structure & -- & Linear(30,64) $\rightarrow$ ReLU $\rightarrow$ Linear(64,64) $\rightarrow$ ReLU $\rightarrow$ Linear(64,1) \\
Output variables & $\mu$, $\sigma^2$ & Mean and variance for Gaussian action sampling \\
Action sampling & -- & $a = \mu + \sqrt{\sigma^2}\,\epsilon,\ \epsilon \sim \mathcal{N}(0,1)$ \\
Sampling time step & \texttt{dt\_new} & $1\times10^{-2}\ \mathrm{s}$ \\
Sampling frequency & \texttt{fs} & $100\ \mathrm{Hz}$ \\
Original data rate & \texttt{dt\_original} & $2\times10^{-4}\ \mathrm{s}$ \\
Bandpass filter & \texttt{bandpass\_filter()} & 4th-order Butterworth (15–55 Hz) \\
Random window length & \texttt{window\_length} & 0.4 s (with random starting points for the windows) \\
\hline
\end{tabular}
\label{tab:training_params}
\vspace{-0.4 cm}
\end{table*}
\begin{figure}[h]
    \centering
\includegraphics[width=0.7\linewidth]{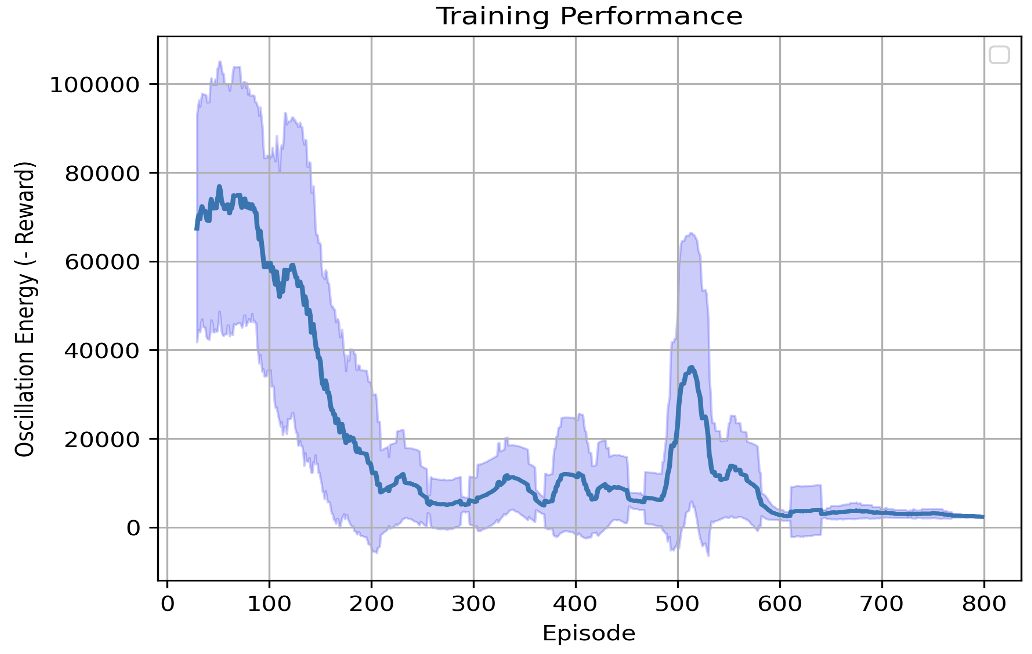}
    \caption{Policy gradient training progress}
    \label{fig:training}
    \centering
    \includegraphics[width=0.7\linewidth]{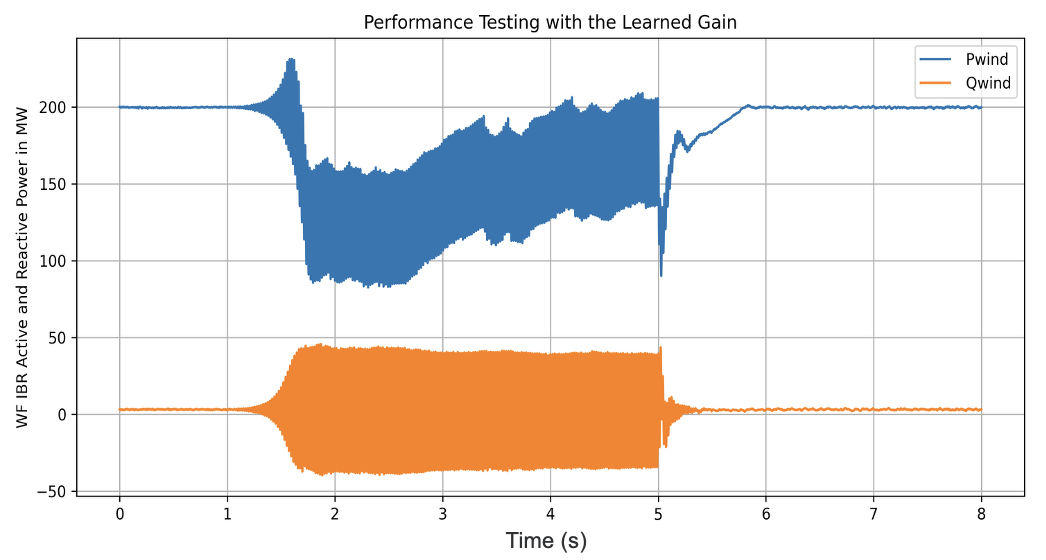}
    \caption{Onset of oscillations due to control gain mistuning at $1$ second, and the learned policy helps to  mitigate the oscillation with optimal control gain at $5$ seconds
}
    \label{fig:performance}
    \vspace{-0.5 cm}
\end{figure}
 
The PSCAD/RL interface has been enabled by the \texttt{mhi.pscad} module functionalities. In the simulated event (Fig.\ref{fig:Currents_WG_Texas}), the proportional gain $K_p$ of both the inner-loop and outer-loop rotor-side controls of the DFIG has been modified to inject SSCI. Fig. \ref{fig:performance} shows that $1$ second onward, system experiences unwanted and dangerous oscillations with sub-synchronous frequency of $48$ Hz. We then conducted the RL policy training as depicted in Algorithm \ref{alg:PG_PSCAD} with parameter settings as given in Table \ref{tab:training_params}, the training performance shown in Fig. \ref{fig:training} with moving window averaging and standard deviations shown with the shaded region.  This shows that the oscillation energy has been decreased over the training episodes, and the optimal SSCI mitigation policy has been learned. Subsequently, we test the performance of such policy by activating the learned control at $5$ seconds in Fig. \ref{fig:performance}. Both the active and reactive power oscillations are significantly minimized when the learned optimal controller is actuated in the system. As the policy network's input is the active power measurements from the grid, $K_p$ of the power control loops are adaptively tuned based on the grid's operating conditions, specifically, in presence of heavy SSCI-induced oscillations. 

\section{Concluding Remarks}
This paper presented an automated IBR gain tuning methodology using simpler deep policy gradients, a class of deep reinforcement learning algorithms, using PSCAD/Python co-simulation interface to mitigate sub-synchronous control interactions. The methodology outlined detailed signal processing and MDP-based learning methodology targeting SSCIs along with considerations for restricted action space explorations to reduce computational complexities. The learning framework has been customized to ingest SSCI-specific grid event data with additional pre-processing modules with down sampling, band-pass filtering, and reward designs. We have conducted numerical tests on a real-world event configuration where complex SSCI interactions occur between DFIG and series capacitor in the network, and shown how adaptive learning-based gain tuning of the IBR control loops can mitigate such oscillations.  Future work will consider developing more efficient algorithms as the computational complexity with the EMT interface is a major challenge for large action spaces with many controllable devices. We will also consider different event settings such as AC/DC hybrid grids and how such adaptive tuning can alleviate concerns with SSOs. 


\subsection*{Acknowledgment}
The authors would like to thank Brett Ross and Quan Nguyen at PNNL for providing helpful suggestions and code snippets regarding the PSCAD's Python API. Authors would like to acknowledge GPT-5 for algorithmic and editing assistance with thorough supervision. 
\bibliographystyle{IEEEtran}
\bibliography{refs}

\end{document}